# Transcript-Free Lightweight Detection of Alzheimer's Disease from Spontaneous Speech Using Handcrafted MFCC-Dominant Acoustic Biomarkers

1st Rashin Gholijani Farahani
Department of computer Engineering,
Ka.c., Islamic Azad University
Karaj,Alborz, Iran
rashin.gholijanifarahani@iau.ir

2nd Azam Bastanfard
Corresponding author
Department of computer Engineering,
Ka.c., Islamic Azad University
Karaj,Alborz ,Iran
bastanfard@iau.ac.ir

**Abstract—It is still hard to find Alzheimer's disease (AD) early, especially when neuroimaging is expensive or tools that depend on language are not available. Spontaneous speech provides a non-invasive signal; however, numerous current methodologies depend on transcripts/ASR or computationally intensive deep models. We offer a simple, audio-only baseline for detecting AD using 176 Cookie Theft recordings from the DementiaBank Pitt corpus (88 AD, 88 controls). WebRTC voice activity detection (VAD) is used to separate speech from non-speech. We take out 99 hand-crafted acoustic-temporal features, including pause and fluency statistics, spectral/prosodic descriptors, and MFCC summaries with Δ and ΔΔ. Evaluation is performed using a stringent speaker-independent GroupShuffleSplit,documenting performance across 30 iterations. A lightweight SVM with an RBF kernel gets an average AUC of 0.674 ± 0.091 across runs. For example, a single split has an AUC of 0.742 and an accuracy of 0.657. We also present an exploratory compact-feature analysis utilizing a Top-20 subset ranked by Random Forest importance; since selection is not nested within training splits, these results may be overly optimistic and are not employed for primary conclusions (AUC 0.719 ± 0.091). The results indicate that transcript-free spectro-temporal and fluency-related cues can facilitate speaker-independent Alzheimer's disease screening from raw audio, establishing a practical foundation for deployment-oriented research.**

**Keywords—**Alzheimer’s disease; spontaneous speech; acoustic biomarkers; pause analysis; WebRTC VAD; MFCC; transcript-free; speaker-independent evaluation; lightweight machine learning; digital screening

## I. Introduction

Alzheimer's disease (AD) is a neurodegenerative disorder that gets worse over time and affects memory, executive function, and communication. Since clinical symptoms frequently manifest only after significant neurodegeneration has transpired, there is a persistent necessity for screening instruments that are timely, accessible, and economical. In this context, spontaneous speech has garnered increasing attention as a non-invasive digital biomarker, as it can indicate early disturbances in temporal control, fluency, and speech planning linked to cognitive decline.

A significant corpus of evidence demonstrates that temporal speech anomalies characterized by prolonged pauses, elevated silence ratios, hesitation patterns, and diminished speaking duration are correlated with Alzheimer's disease (AD) and mild cognitive impairment (MCI) [1]–[6]. Cross-lingual studies in English, Chinese, Swedish, and Spanish indicate that cognitively impaired speakers display pauses that are longer, more frequent, and more variable compared to healthy controls [1], [2], [5], [7]. Additionally, the distributional characteristics of pauses, such as upper-tail statistics (e.g., P90), long-to-short pause ratios, and decreased articulation time, have been identified as useful for differentiating pathological from normative aging [3, 4, 7].

In addition to timing-related markers, acoustic and spectral cues can offer supplementary diagnostic information. Alterations in vocal energy, rhythm stability, spectral dynamics, and MFCC-based representations have been associated with early cognitive decline [10]–[12]. Recent advancements in deep learning, encompassing convolutional networks, self-supervised speech encoders (e.g., wav2vec-style models), and multimodal pipelines, have demonstrated exceptional efficacy on benchmark datasets such as ADReSS and ADReSSo [7], [12], [14], [15]. But a lot of these systems need big training sets, transcript-based language models (often made with ASR), and lots of GPU power, which can make them harder to understand and less useful in real-world or low-resource screening situations.
Even though things are moving quickly, there are still some problems that need to be fixed. First, a lot of the best new methods depend on transcript-based linguistic features, which means they are affected by differences in language, dialect, and ASR errors [6], [16]. Second, deep neural architectures can require a lot of processing power, which can make it hard to use them regularly or for large-scale screening. Third, evaluation protocols don't always make sure that speakers are independent, and leaking speaker identity can make performance estimates look better than they are, especially on smaller datasets [1], [16], [17]. These factors drive the need for lightweight, reproducible, and leakage-aware baselines that work directly with raw audio.

To deal with these problems, we present a lightweight, transcript-free, and fully audio-based baseline for AD detection. We use 176 spontaneous speech recordings from the DementiaBank Pitt corpus (Cookie Theft task) and a strict speaker-independent evaluation protocol to get 99 handcrafted temporal–acoustic features. These features include pause and fluency statistics, spectral/prosodic descriptors, and MFCC summaries with Δ and ΔΔ. A straightforward SVM utilizing an RBF kernel serves as a computationally efficient benchmark, attaining an average AUC of 0.674 ± 0.091 across 30 repeated speaker-independent splits; we present a single held-out split solely for demonstration (AUC 0.742). In accordance with recent clinical-AI guidelines advocating for accessible, interpretable, and scalable digital screening instruments for cognitive evaluation [16], [17], our endeavor seeks to create a transparent baseline that facilitates deployment-focused research in authentic evaluation settings.

## II. Related Works

Speech-based detection of Alzheimer's disease (AD) has been investigated through various complementary approaches, including pause-derived temporal analysis, acoustic–prosodic modeling, MFCC and other spectral representations, transcript-based transformer models, end-to-end deep learning systems, and systematic reviews. These lines of work together make a case for lightweight and language-independent methods, like the framework that was created in this study.

Pausing behavior is frequently cited as a significant indicator of cognitive impairment. Systematic reviews show that people with AD and MCI often have longer, more frequent, and more variable pauses, as well as higher silence ratios and lower articulation rates [1], [2]. Empirical research utilizing multilingual datasets comprising English, Chinese, Swedish, and Spanish indicates that pause duration statistics, distributional characteristics (such as skewness and kurtosis), and long-to-short pause ratios can facilitate the differentiation between cognitively impaired and healthy speakers [3]–[5].

Numerous studies investigate the alterations in pause behavior corresponding to the severity of disease. Sluis et al. documented escalating long silent pauses (>2 s) among healthy aging, mild dementia, and moderate dementia cohorts [6]. Liu et al. suggested effective, language-agnostic pipelines for delineating pause boundaries and underscored their applicability in spontaneous speech tasks [7]. Recent research emphasizes contextual and modeling factors: Khatri et al. demonstrated that pause distributions, when combined with contextual cues, can enhance Alzheimer's disease (AD) classification [8]. Additionally, Pu and Zhang incorporated pause durations into BERT positional encodings, achieving an accuracy of 83.1% on ADReSSo [31]. Soleimani et al. showed that using contrastive learning to encode pauses and filler words (like "uh" and "um") can make performance better with fewer parameters [32]. Adversarial studies indicate that altering pause durations may influence classifier decisions, highlighting sensitivity to temporal gaps in speech [10]. Multi-cohort analyses indicate that pause variability, silence ratio, and long-tail pause distributions continue to provide valuable insights across diverse recording conditions [34].

He et al. [38] suggested an automated, language-agnostic bimodal pause analysis employing log-normal distribution fitting on connected speech, establishing an approximate threshold (~180 ms) that differentiates short (phonetic) pauses from long (cognitive) pauses. This is generally in line with common practical thresholds in the range of 180 to 200 ms, which backs up the threshold choices used in transcript-free pipelines. In sum, these studies support the use of pause-derived temporal biomarkers as understandable signs of cognitive decline.

In addition to temporal pauses, acoustic–prosodic cues offer supplementary diagnostic significance. Previous research indicates that voice quality and stability metrics including spectral descriptors, energy dynamics, and associated prosodic statistics may exhibit alterations in Alzheimer's Disease (AD) and notably in Mild Cognitive Impairment (MCI) [11]. Extensive clinical studies utilizing paralinguistic acoustic features from brief spontaneous speech have demonstrated significant predictive efficacy for dementia and mild cognitive impairment (MCI), along with correlations across various cognitive domains [28]. Themistocleous et al. demonstrated that voice instability and diminished fluency can differentiate MCI from healthy aging in the absence of linguistic indicators [12]. García-Gutiérrez et al. [39] employed eGeMAPS features to forecast amyloid status in mild cognitive impairment (MCI) from spontaneous speech, achieving an area under the curve (AUC) of 0.79 and correlating acoustic markers with Alzheimer's disease-related pathology. Explainable-ML studies also stress the importance of energy fluctuations and spectral dynamics. For example, Oiza-Zapata and Gallardo-Antolín used SHAP to show how important features like spectral flux and amplitude variability are. They also reported good results on ADReSS with an XGBoost pipeline and used data-quality filters [13]. These results prompt the incorporation of lightweight descriptors, including ZCR, spectral centroid, bandwidth, and flux, in the current study.

MFCCs are still widely used to find AD because they are sensitive to how well someone articulates and how stable their voice is. Numerous studies employing classical machine learning and CNN-based pipelines validate the discriminative efficacy of MFCC, Δ, and ΔΔ trajectories for the detection of Alzheimer's Disease and Mild Cognitive Impairment [11], [14]. Ahn et al. utilized MFCC-related spectral representations in conjunction with DenseNet121, achieving high screening accuracy in elderly Korean speech [14]. Vats et al. examined high-resolution spectral features, including SFCC, to detect subtle spectro-temporal distortions associated with cognitive decline [15]. These results validate MFCC+Δ+ΔΔ as a lightweight yet effective representation for transcript-free Alzheimer's disease detection. More proof from large-scale explainable frameworks shows that spectral irregularities and temporal-prosodic instability have a big effect even when there is a lot of noise in the real world [35].

Another important area of research uses ASR and transformer encoders to extract linguistic features from transcripts. Ding et al. and Qi et al. found that transformer-based methods often do very well on ADReSS/ADReSSo [16], [17]. However, systematic reviews consistently underscore the limitations of transcript-dependent pipelines, notably their susceptibility to ASR errors, linguistic structure, and dialectal variation [1], [16]. Research on linguistic micro-features indicates that decline influences coherence and grammatical structures; however, these indicators may also vary based on language proficiency and recording context [18]. These worries push for fully audio-based options for language-independent deployment.

End-to-end deep learning models, such as CNNs, LSTMs, self-supervised encoders, and multimodal architectures, have shown strong performance on benchmark tests. Ahn et al. illustrated spectrogram-based CNN pipelines for screening [14]. To lower the cost of transformer calculations, compact, attention-efficient designs like HAFFormer have been looked into for ADReSS-M. These designs have a smaller model size and work better [33]. Rezaii et al. put forward a combination of YOLO and LLM for analyzing digital voice in the LEADS dataset, and they found that the AUC values for the MCI and EOAD tasks were very high [19]. Multimodal fusion models that combine speech with imaging and clinical variables have also done well on datasets like ADNI and NACC [20], [21]. Even though these systems work well, they often need a lot of data and processing power. Reviews stress the need for lighter, easier-to-understand, and more clinically scalable options [1], [16], [17]. Recent research indicates that transformer-based encoders may surpass MFCC-only pipelines in multilingual contexts, yet they remain susceptible to ASR and language variation [36], thereby reinforcing the necessity for the current transcript-free pipeline.

Imaging-based methods, such as MRI/DTI and fusion networks, have shown very high accuracy in controlled settings. Recent imaging-fusion and hybrid deep architectures demonstrate accuracies in the mid-to-high 90s on clinical datasets [20], [21]. Imaging, on the other hand, needs special tools and clinical workflows, which makes speech-based screening easier to use in places that are far away or don't have a lot of resources.
Systematic reviews consistently highlight significant obstacles to speech-based Alzheimer's disease detection, such as inadequate standardization, insufficient dataset sizes, generalization deficiencies, privacy issues, and inconsistent evaluation protocols [1], [6], [16], [17]. Reviews also stress the need for tools that are clear, cheap, and have been tested in a clinical setting to help with real-world use [22]. In response, this study aims for a baseline that is language-independent, interpretable, and aware of leaks, with strict evaluation that doesn't depend on the speaker. The availability of standardized speech corpora is a fundamental prerequisite for developing reliable speech-based diagnostic systems. In low-resource languages such as Persian, the AVA audio-visual corpus provides a structured benchmark for speech research and highlights the importance of creating language-specific resources for speech analysis and machine learning applications [40].

## III. METHODOLOGY

This study introduces a lightweight, language-independent, and entirely audio-based pipeline for the early detection of Alzheimer's disease (AD) from spontaneous speech. The pipeline includes: (1) preparing the dataset with strict speaker-independent evaluation, (2) standardizing audio preprocessing and using WebRTC VAD to separate speech from non-speech, (3) manually extracting acoustic-temporal features, and (4) training and testing lightweight classifiers on repeated speaker-independent splits.

### A. Dataset and Splitting That Doesn't Depend on the Speaker

We utilize 176 Cookie Theft picture-description recordings from the DementiaBank Pitt corpus [30], evenly distributed across classifications (88 AD, 88 healthy controls (HC)). For each recording, there is a participant identifier that was taken from the filename by looking at the part of the name before the first hyphen ("-"), which is the speaker (group) label.

To avoid leaking speaker identities, all experiments require strict speaker independence. This means that speakers in the test set are not seen at all during training. We use GroupShuffleSplit (test_size = 0.2) and group by participant_id. Results are presented in two complementary contexts: (i) an illustrative split (random_state = 42), utilized for a specific classification report; and (ii) consistent performance across 30 repeated speaker-independent splits (n_splits = 30, random_state = 42), expressed as mean ± standard deviation.

### B. Audio Preprocessing and Segmentation Based on VAD

All recordings are standardized to minimize variability not associated with cognitive status and to guarantee uniform temporal and spectral feature extraction:

Loading and sampling again. Librosa loads the audio, changes it to mono, and resamples it to 16 kHz.

Filtering by length. To keep the statistics stable, recordings shorter than 0.3 seconds are not included.

Normalizing amplitude. To normalize the peak, you divide the waveform by its highest absolute amplitude.

WebRTC voice activity detection (VAD). WebRTC VAD is used to separate speech from non-speech by setting the aggressiveness level to 2, the frame length to 30 ms, and the hop size to 30 ms (non-overlapping frames). To get a binary speech mask, the normalized waveform is changed to 16-bit PCM and then processed frame by frame.

Pause extraction without a transcript. Consecutive frames that don't have speech are put together to make candidate pauses. A pause is only counted if it lasts longer than 0.2 seconds (the pause threshold). There is no extra "merge gap" post-processing done after the framewise accumulation.

There is no need for ASR, lexical content, or language-specific resources with this segmentation strategy.

### C. Handcrafted Feature Extraction

For each recording, we make a small set of custom-made acoustic-temporal descriptors that are meant to capture timing, fluency, and spectral characteristics. To help with feature construction, auxiliary timing variables like total duration and estimated speaking time are calculated. The final model input, on the other hand, is made up of 99 acoustic features, which are grouped into three groups:

Pause-based temporal features (13). There are a lot of pauses, the total length of the pauses, the silence ratio, and summary statistics of the lengths of the pauses (mean, standard deviation, median, maximum, 90th percentile, skewness, and kurtosis). We also calculate the percentages of short and long pauses using 0.3 s and 1.0 s, respectively, as the thresholds.

Acoustic and prosodic features (8). The average and standard deviation of the zero-crossing rate, spectral centroid, spectral bandwidth, and spectral flux.
Features based on MFCCs (78). MFCCs (13) and first- and second-order deltas (Δ and ΔΔ) make up 39 coefficients. There are 39 × 2 = 78 features because each coefficient is summarized by its mean and standard deviation.

Table I — Summary of Handcrafted Feature Categories (Total = 99)

| **Feature group** | **Dim** | **Included descriptors (as implemented)** |
|---|---|---|
| Pause-based temporal | 13 | n_pauses, total_pause, silence_ratio; pause stats: pause_mean, pause_std, pause_median, pause_max, pause_p90, pause_skew, pause_kurt; ratios: short_pause_ratio (pause < 0.3s), long_pause_ratio (pause > 1.0s) |
| Acoustic–prosodic | 8 | Mean & std of ZCR, spectral centroid, spectral bandwidth, spectral flux |
| MFCC + Δ + ΔΔ | 78 | 13 MFCC + 13 Δ + 13 ΔΔ = 39 coeffs; each summarized by mean and std → 39×2 |
| **Total** | **99** | Combined pause-temporal + acoustic–prosodic + MFCC-based descriptors |

D. Classification Models and Training Pipeline

We evaluate lightweight classifiers suitable for small datasets:

- **SVM (RBF kernel):** SVC(kernel="rbf", probability=True, class_weight="balanced"), with scikit-learn defaults C = 1.0 and gamma = "scale".
- **Random Forest:** n_estimators = 400, class_weight = "balanced", random_state = 0.
- **XGBoost (optional):** n_estimators = 300, max_depth = 4, learning_rate = 0.05, subsample = 0.9, colsample_bytree = 0.9, tree_method = "hist".

All models are trained using a scikit-learn **Pipeline** to avoid preprocessing leakage across speakers. Within each split, imputation and scaling are fit on training speakers only and then applied to held-out speakers:

- SimpleImputer(strategy="mean")
- StandardScaler()
- classifier

E. Evaluation Metrics and Reporting

We perform speaker-independent evaluation using GroupShuffleSplit. Performance is reported using:

- **Accuracy**
- **F1-score** (binary F1, with **AD treated as the positive class**)
- **ROC-AUC**

We report: (i) one illustrative split (random_state = 42), and (ii) mean ± std over **30 repeated** speaker-independent splits (n_splits = 30).

F. Exploratory Feature-Subset Analysis (Top-20)

We additionally conduct a compact-feature experiment by ranking features using Random Forest feature importance and evaluating models on a selected **Top-20** subset. In the current implementation, feature importance is computed **once on the full dataset** and the same Top-20 subset is reused across splits. Because selection is **not nested within each training split**, this analysis may introduce optimistic bias and is therefore treated as **exploratory** rather than a primary result. A leakage-safe alternative would recompute feature ranking using training speakers only within each split and then evaluate on that split's held-out speakers.

# IV. Results

This section presents: (i) a descriptive analysis of pause-derived temporal markers; (ii) qualitative complementarity between temporal and spectral (MFCC/prosodic) cues; (iii) performance on an illustrative speaker-independent split; (iv) robustness under repeated speaker-independent evaluation; and (v) exploratory compact-subset and feature-group analyses.

A. The makeup of the dataset and the pause distributions

After preprocessing, the dataset consists of 176 Cookie Theft recordings from the DementiaBank Pitt corpus (88 AD, 88 HC). All recordings are resampled to 16 kHz mono, peak-normalized, and split into speech and non-speech intervals using WebRTC VAD (level = 2; 30 ms non-overlapping frames). This creates a pipeline that is completely based on audio and does not depend on language.

Fig 1 shows the actual distribution of the average pause length for AD and HC. Both distributions are right-skewed. In general, AD recordings show a shift toward longer pauses and a heavier upper tail. Fig. 2 shows the distribution of the 90th-percentile pause duration (pause_p90), which also shows that AD has bigger upper-tail values and more extreme outliers. These descriptive trends align with previous research associating heightened pausing and temporal irregularities with cognitive decline [9]–[13]. We stress, though, that descriptive separation doesn't always lead to stable predictive performance that doesn't depend on the speaker (see Section IV-F).

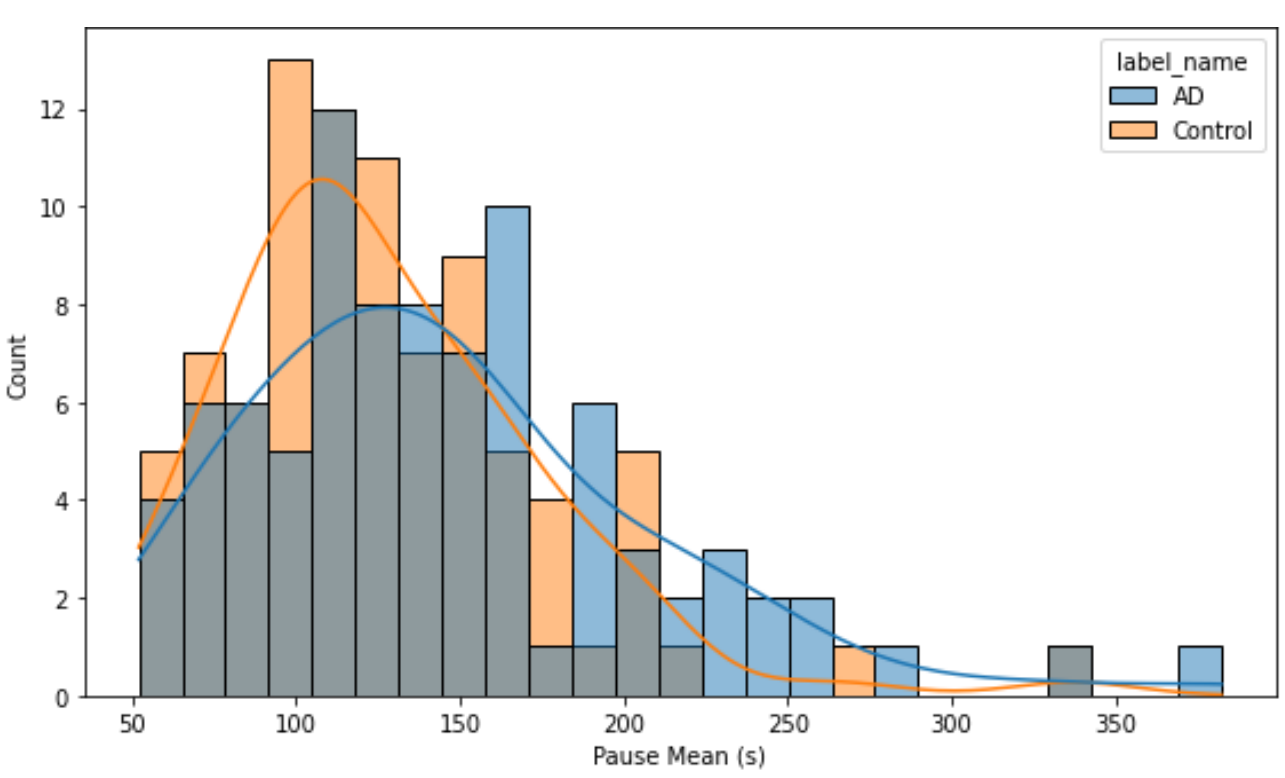

Fig. 1 Empirical distribution of mean pause duration for AD and HC recordings.

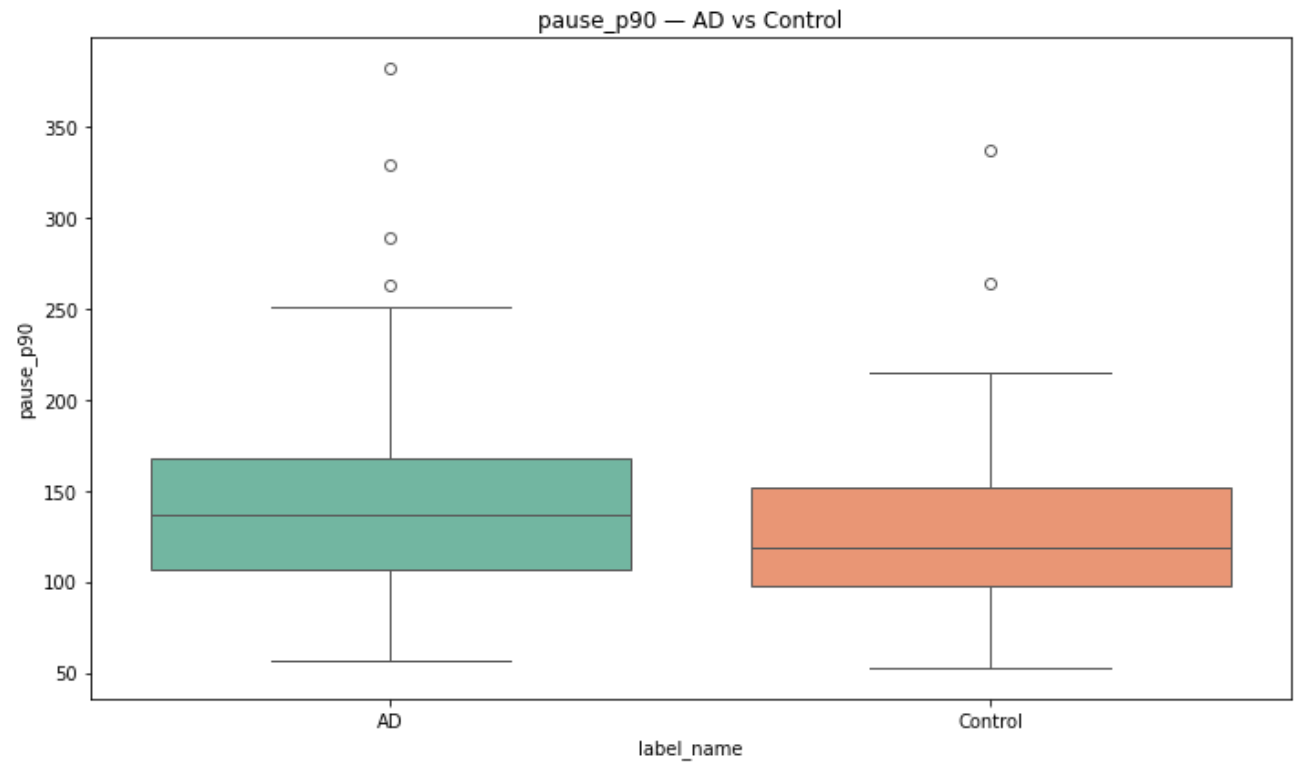

Fig. 2 Distribution of the 90th-percentile pause duration (pause_p90).

## B. Joint Behavior of Pause and MFCC Features

To show how temporal and spectral cues might work together, Fig. 3 shows the joint distribution of mean pause duration and mean MFCC-1. Qualitatively, AD samples are more prevalent at extended pause durations and, for similar pause durations, typically demonstrate reduced MFCC-1 values. Although these patterns are observational, they encourage the integration of pause-derived markers with MFCC/prosodic descriptors into a cohesive representation [10], [14], [16], [22].

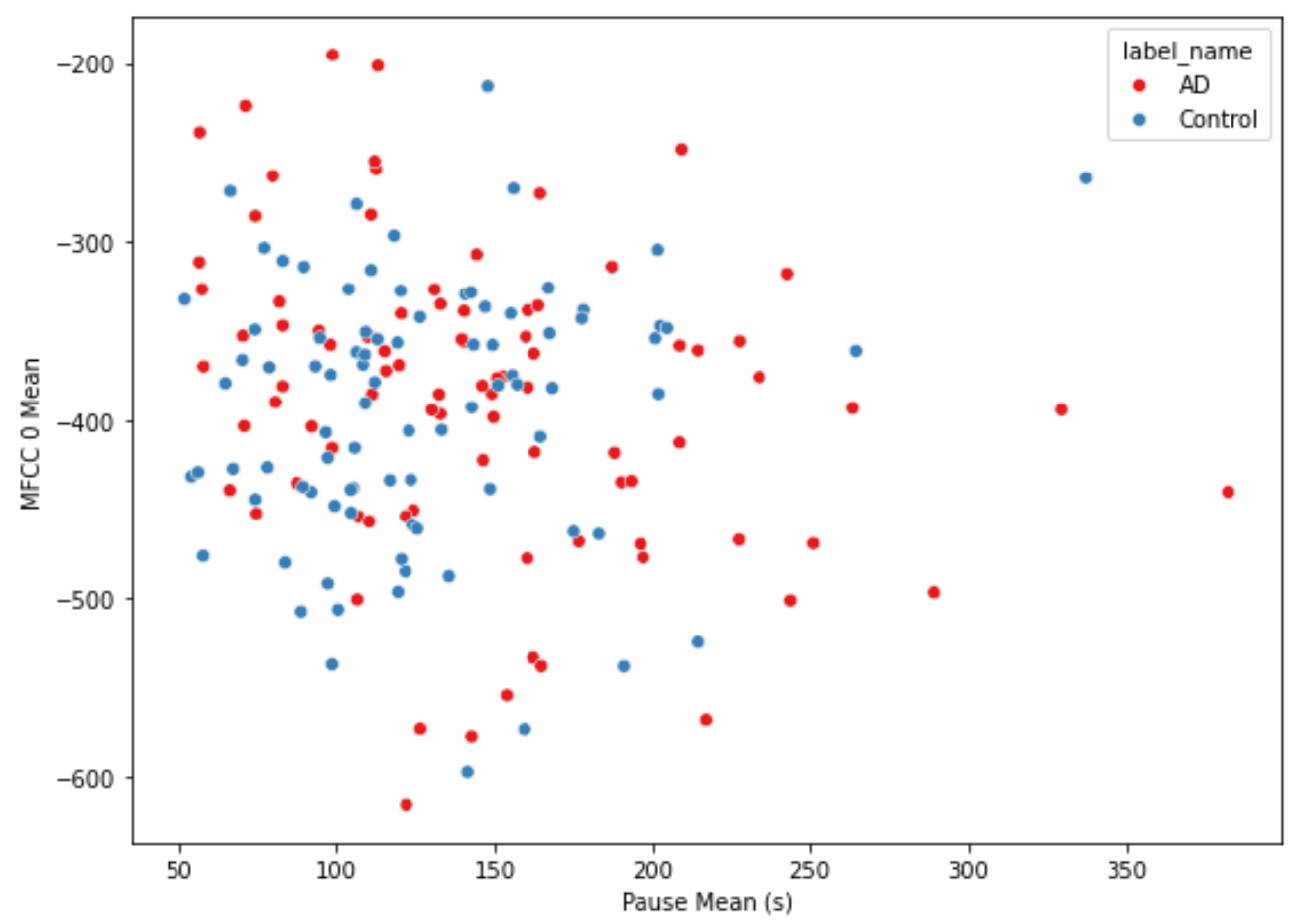

Fig. 3 Joint distribution of mean pause duration and mean MFCC-1.

## C. Illustrative Speaker-Independent Classification (Single Split)

The SVM classifier gets this score on one held-out speaker partition when it uses the full 99-feature representation and a strict speaker-independent split (GroupShuffleSplit, test_size = 0.2):

- Accuracy: **0.657**
- F1 (AD): **0.667**
- AUC: **0.742**

Fig. 4's ROC curve shows separation above chance, which means that purely acoustic temporal and spectral cues (without any words) can help tell the difference between speakers when strict speaker independence is needed. These single-split values are provided solely for illustrative purposes; the principal conclusions are based on repeated speaker-independent evaluations (Section IV-D).

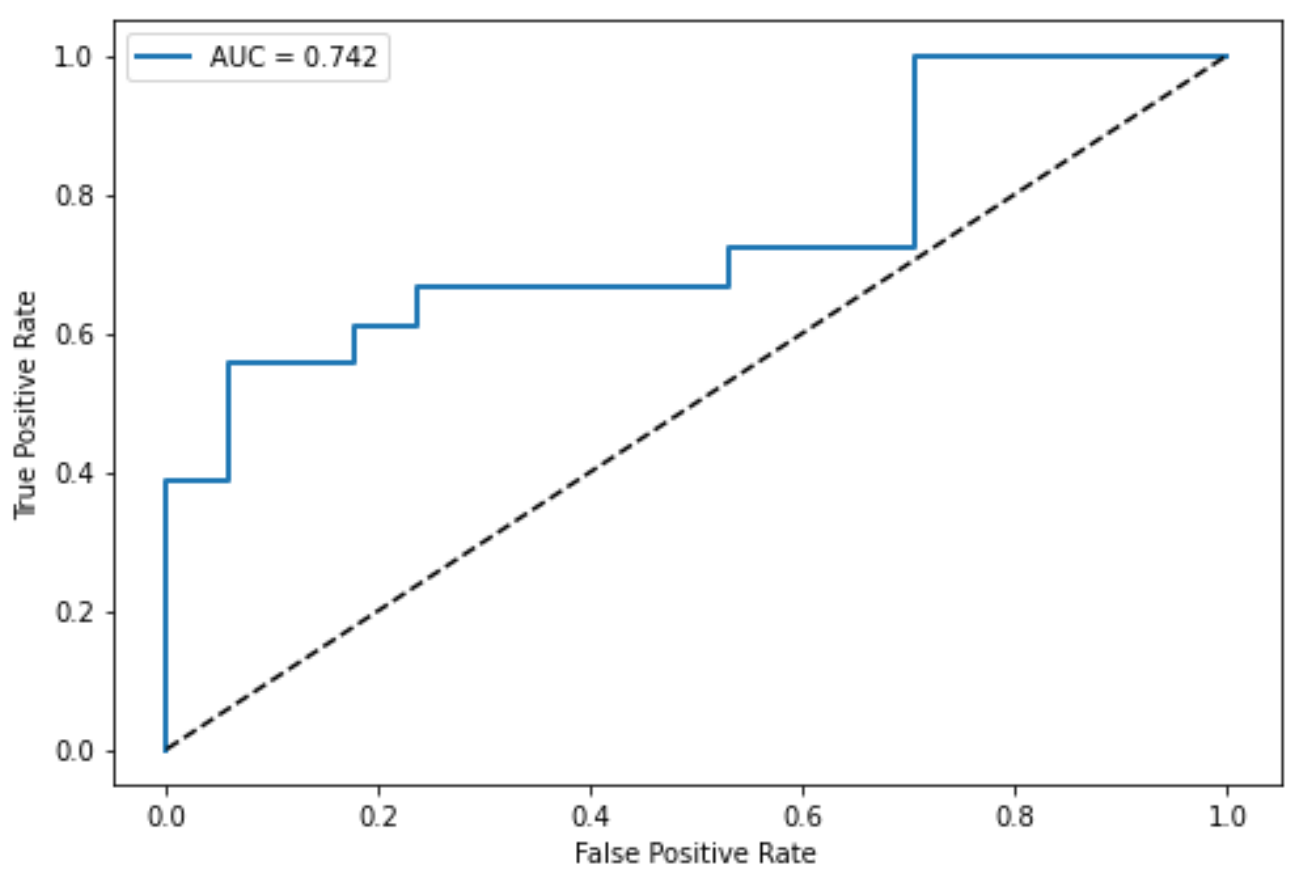

Fig. 4 ROC curve of the SVM classifier on a representative speaker-independent split.

D. Benchmarking Lightweight Classifiers (30 Speaker-Independent Splits)

To test robustness beyond a single partition, we run SVM, Random Forest, and XGBoost 30 times in a row, with about 20% of the speakers held out each time. In each run, imputation and scaling are trained on speakers that are not being tested and then used on speakers that are being tested. Table II shows the mean and standard deviation for Accuracy, F1 (AD), and ROC-AUC.

Table II — Performance of Lightweight Classifiers Across 30 Speaker-Independent Splits

| **Model** | **Accuracy (mean±std)** | **F1(AD) (mean±std)** | **AUC (mean±std)** |
|---|---|---|---|
| SVM (RBF) | 0.614 ± 0.077 | 0.600 ± 0.090 | 0.674 ± 0.091 |
| Random Forest | 0.584 ± 0.083 | 0.567 ± 0.102 | 0.651 ± 0.095 |
| XGBoost | 0.575 ± 0.069 | 0.581 ± 0.075 | 0.622 ± 0.081 |

SVM has the highest mean AUC across all runs, which means it is the most reliable of the lightweight baselines that were tested. The single-split AUC (0.742) is in the middle of the range of values seen in repeated speaker-independent runs. As a result, the interpretations below are based on the repeated-split results.

E. Feature-Importance and Compact 20-Feature Subset

We use Random Forest feature importance to find the Top 20 features and then look for compact representations. The chosen subset has mostly MFCC/Δ/ΔΔ statistics and a few spectral/prosodic stability cues, such as flux_std, zcr_std, and spectral bandwidth/centroid summaries. This is in line with the ranking.
This feature selection is calculated once on the entire dataset (i.e., not nested within each training split), so the compact-subset experiment may introduce optimistic bias and is therefore considered exploratory. The main findings of this research are derived from the comprehensive 99-feature assessment presented in Section IV-D. In this exploratory context, retraining on the Top-20 subset produces:

- **SVM AUC:** 0.719 ± 0.091
- **RF AUC:** 0.713 ± 0.094

These results suggest that a compact spectro-temporal subset *may* retain a substantial fraction of the discriminative signal; however, leakage-safe confirmation would require nested selection within each training split.

F. Feature-Group Analysis

We train models using (i) pause-only temporal descriptors, (ii) acoustic–prosodic descriptors, (iii) MFCC-based descriptors, and (iv) the full 99-feature set, all under the same 30-run speaker-independent protocol, to see how each feature family adds to the model. Table III shows what each group is, and Table IV shows how well they did.

Table III — Feature Group Definitions Used in the Analysis

| **Feature group** | **Dim** | **Description** |
|---|---|---|
| Pause-only | 13 | Pause duration statistics and temporal fluency markers derived from VAD segmentation |
| Acoustic–prosodic only | 8 | ZCR and spectral centroid/bandwidth/flux summaries |
| MFCC-only | 78 | MFCC + Δ + ΔΔ summaries (mean and std) |
| Full feature set | 99 | Combined pause + acoustic–prosodic + MFCC descriptors |

Table IV— Performance of feature groups under 30 repeated speaker-independent splits

| **Feature group** | **Dim** | **Acc (mean±std)** | **F1(AD) (mean±std)** | **AUC (mean±std)** |
|---|---|---|---|---|
| Pause-only | 13 | 0.529 ± 0.056 | 0.481 ± 0.105 | 0.432 ± 0.065 |
| Acoustic–prosodic | 8 | 0.589 ± 0.059 | 0.574 ± 0.093 | 0.644 ± 0.092 |
| MFCC-only | 78 | 0.610 ± 0.075 | 0.603 ± 0.088 | 0.654 ± 0.098 |
| Full-99 | 99 | 0.614 ± 0.077 | 0.600 ± 0.090 | 0.674 ± 0.091 |

Pause-only markers show descriptive patterns in the pause distribution plots, but they don't provide stable standalone discrimination when evaluated strictly by speaker-independent standards in this context. This difference shows that descriptive separation at the group level may not lead to strong predictive performance, and that features derived from pauses may be sensitive to how segmentation works and how different speakers sound. MFCC-based descriptors, on the other hand, offer stronger and more stable discrimination. Adding a few prosodic stability cues to them is in line with what was seen in the exploratory Top-20 subset.

G. Comparison with Previous Research

Table V shows the main findings of earlier studies on using speech to find AD. Different studies report different metrics, and the results can't be directly compared because the datasets, supervision (transcripts vs. audio-only), and evaluation protocols (speaker-dependent vs. speaker-independent) are all different. So, the table is meant to put our baseline in context, not to say that it is better.

Table V — Contextual Comparison With Prior Speech-Based AD Detection Studies

| **Study** | **Modality** | **Model** | **Dataset** | **Reported Performance** | **Key Difference** |
|---|---|---|---|---|---|

| Pu & Zhang (2025) | Pause + BERT | Transformer | ADReSSo | Acc ≈ 83.1% | Requires transcripts/LM + higher compute |
|---|---|---|---|---|---|
| Khatri et al. (2024) | Pause distrib. | Classical ML | DementiaBank | F1 ≈ 0.78 | Different settings/protocol |
| García-Gutiérrez (2024) | Acoustic-only | Ensemble | Clinical (ES) | F1 ≈ 0.92 | Larger clinical dataset |
| Liu et al. (2023) | Pause-only | Classical ML | DementiaBank | Acc ≈ 0.70 | Temporal-only baseline |
| **This work** | Pause + Acoustic + MFCC | SVM (RBF) | DementiaBank (176) | **AUC = 0.674 ± 0.091 (Full-99, repeated)**; **0.719 ± 0.091 (Top-20, exploratory)** | Lightweight, interpretable, CPU-only, speaker-independent |

H. summary of the most important results

Descriptive analyses show that AD recordings usually have longer pauses and higher upper-tail pause statistics (like P90), which is in line with what has been found before.

When evaluated strictly without regard to the speaker, spectro-temporal MFCC features give the most stable standalone performance. In this case, pause-only markers are not a reliable way to tell the difference.

The SVM with the full 99-feature set gets an AUC of 0.674 ± 0.091 over 30 repeated speaker-independent splits. This is better than other lightweight baselines that were tested in terms of mean AUC.

An exploratory (non-nested) analysis of a small Top-20 subset shows an AUC of about 0.719 ± 0.091. This suggests that dimensionality reduction is possible, but it needs to be confirmed with a selection method that doesn't let any data leak.

In general, these results show that it is possible to use transcript-free, lightweight, and speaker-independent acoustic baselines to find AD in raw audio. However, they also show that pause-only markers may not be enough as stand-alone features when strict testing is done.

## V. Discussion

This study examined the early detection of Alzheimer's disease (AD) from spontaneous speech utilizing a transcript-free, entirely audio-driven pipeline founded on handcrafted acoustic-temporal biomarkers. The complete 99-feature SVM baseline attains an average AUC of 0.674 ± 0.091 across 30 iterations of GroupShuffleSplit in a stringent speaker-independent assessment. For demonstration purposes, only one held-out split is shown (AUC 0.742). We also did an exploratory compact-feature experiment (Top-20) that gave us a higher mean AUC (0.719 ± 0.091) while greatly reducing dimensionality. However, since feature ranking is not nested within training splits, these compact-subset results are seen as exploratory and are not used to draw main conclusions.

A. Understanding Temporal and Spectro-Temporal Biomarkers

Descriptive analyses indicate that AD recordings typically display extended pauses and elevated upper-tail pause statistics (e.g., P90), aligning with previous research linking temporal dysfluency to cognitive-linguistic disruption [5], [7], [9], [10]. The simultaneous visualization of pause metrics alongside MFCC-derived cues suggests that temporal and spectral characteristics may convey complementary information. Qualitatively, AD samples are more prevalent in areas characterized by longer pauses and lower MFCC-1 values, potentially indicating a blend of disrupted planning and modified articulatory/spectral stability [11, 12, 28].

Repeated speaker-independent experiments have shown that pause-only features are not reliable as standalone discriminators in this context (AUC below chance). This emphasizes a fundamental difference between group-level descriptive separation and strong speaker-independent predictability. In pipelines without transcripts, pause statistics can be affected by how segmentation works (like VAD behavior), how recordings vary, and how speakers differ in their non-linguistic silence patterns. On the other hand, MFCC-based spectro-temporal descriptors offer more stable discrimination. The exploratory Top-20 ranking indicates that a limited number of spectral/prosodic stability cues (e.g., spectral flux and ZCR variability) may also play a role. In general, the repeated tests show that spectro-temporal patterns, not just pause statistics, are what really drive strong discrimination.

B. Positioning in relation to current speech-based AD systems

Recent transformer-based and multimodal methodologies frequently demonstrate superior performance on benchmark datasets [14]–[16], [31]–[33]. But a lot of these kinds of systems depend on linguistic representations that depend on transcripts, content that comes from ASR, or deep architectures that use a lot of computing power. These options can lead to language dependence and transcription bias, and they may not work as well in low-resource or real-world screening situations [1]–[3], [6], [16], [17], [24].

In contrast, this work focuses on deployment-oriented constraints: (i) operation that is fully audio-only and does not require a transcript, (ii) handcrafted biomarkers that are lightweight and easy to understand, and (iii) strict speaker-independent evaluation to prevent identity leakage. In these conservative conditions and with a small number of subjects, the performance we saw is what we would expect from lightweight acoustic baselines. The aim is not to optimize benchmark accuracy, but to create a clear and replicable standard for realistic transcript-free screening research.

C. Real-World Effects and Possible Uses

Multimodal clinical pipelines that combine neuroimaging and clinical variables can give more accurate diagnoses in controlled settings, but they are expensive and not always practical for large-scale or remote screening. Speech-only methods are better understood as cost-effective triage instruments that may identify individuals for subsequent neuropsychological or clinical evaluation, especially when privacy-preserving, language-neutral solutions are preferred [2], [4], [17], [22], [28]. The proposed pipeline functions without transcripts, suggesting theoretical applicability across languages; however, this assertion necessitates cross-corpus validation and fairness assessments among demographic groups.

D. Limitations

Several limitations should be considered when interpreting these findings:

1. **Dataset size and task specificity.** Results are based on 176 recordings from a single elicitation task (Cookie Theft), which limits generalizability.

2. **Task and interaction dependence.** Speech characteristics may vary across tasks, interview styles, and conversational contexts; ecological validity requires multi-task and cross-corpus evaluation.

3. Demographic and clinical confounders. Factors such as age, education, comorbidities, and medication can influence speech; future work should incorporate demographic adjustment and subgroup/fairness analyses where possible.

4. Recording and segmentation sensitivity. Transcript-free pipelines may be affected by recording conditions and segmentation behavior; more robust processing (e.g., calibrated VAD, diarization-aware segmentation) could improve stability, especially for pause-derived features.

5. Binary classification setting. Clinically, modeling MCI and progression trajectories (HC→MCI→AD) is highly relevant; larger cohorts are required to reliably study this continuum.

6. Exploratory compact-subset analysis. The Top-20 improvement should be interpreted cautiously because feature ranking is not nested within training splits; leakage-safe confirmation would require per-split ranking using training speakers only.

More broadly, comparisons to previously reported results must be interpreted carefully because speech-based performance on limited datasets can be strongly influenced by evaluation protocols, particularly speaker overlap. Strict speaker-independent testing yields a more conservative but realistic estimate of generalization.

E. Optional Exploratory MCI Analysis

If included, this subsection should be explicitly marked as exploratory due to limited sample size. Preliminary experiments on the available MCI subset (≈33 samples) suggest intermediate behavior between HC and AD, with an MCI-vs-HC AUC of approximately 0.68 under speaker-independent evaluation. These results are suggestive rather than conclusive and motivate future investigation with larger MCI cohorts.

F. Outlook and Future Directions

Overall, the results indicate that informative AD-related cues can be extracted from raw audio using a lightweight, transcript-free pipeline under rigorous speaker-independent evaluation. Promising future directions include: (i) improving the robustness of pause extraction (e.g., calibrated VAD and diarization-aware segmentation), (ii) incorporating temporal context modeling without transcripts, (iii) cross-task and cross-language transfer evaluation, and (iv) external validation on independent cohorts. While handcrafted audio-only baselines are not intended to replace multimodal clinical systems, they provide an interpretable and deployable foundation for scalable screening-oriented research.

## VI. Conclusion

It is still hard to find Alzheimer's disease (AD) early on, especially in places where neuroimaging is expensive or language-dependent tools are hard to use. This study illustrates that Alzheimer's disease-related disparities can be identified through meticulously crafted acoustic-temporal biomarkers derived directly from spontaneous speech, circumventing the need for transcripts, linguistic embeddings, or resource-intensive deep learning architectures.

Our framework, which doesn't use transcripts and only uses audio, combines VAD-derived pause/fluency statistics with MFCC-based spectro-temporal descriptors to find acoustic patterns that could be related to cognitive and articulatory decline. When tested strictly without regard to the speaker, an SVM with the full 99-feature representation gets an average AUC of 0.674 ± 0.091 over 30 repeated speaker-independent splits. For demonstration purposes, only one held-out split is shown (AUC 0.742). These results show that the evaluation protocol is conservative and that the research can be used as a clear baseline for deployment-oriented research.

We also looked at a compact feature-subset setting that used a Top-20 selection ranked by Random Forest importance. This gave us an AUC of 0.719 ± 0.091 and cut down on the number of dimensions. This subset result is considered exploratory because feature ranking was not nested within each training split, and it is not used for primary conclusions. To ensure leakage-safe confirmation, selection must be done per split using only training speakers.

It is important to recognize a number of limitations. The study is limited by its small sample size, a single elicitation task (Cookie Theft), and binary labeling; it also does not include demographic or clinical covariates that could influence speech. Additionally, DementiaBank recordings may differ in acquisition conditions, such as background noise, microphone handling, and encoding. As Gauder et al. [37] pointed out, non-speech or recording-related artifacts can sometimes affect above-chance classification performance. This shows how important it is to be aware of artifacts and control for confounding factors in speech-based AD research.

Subsequent research ought to authenticate the methodology on more extensive and diverse cohorts, broaden the assessment to various tasks and languages, integrate artifact-aware and domain-robust processing, and progress towards multi-class and longitudinal modeling in accordance with biomarker-informed diagnostic frameworks. This study offers a clear and resource-efficient foundation for transcript-free speech-based Alzheimer's disease screening and underscores the potential significance of spectro-temporal acoustic cues in realistic speaker-independent assessments.

## Acknowledgment

The authors gratefully acknowledge the use of the DementiaBank Pitt Corpus, collected and maintained with support from the U.S. National Institute on Aging (NIA) under grants AG03705 and AG05133. The clinical cohort is derived from Becker et al. (1994) [30].

## References

[1] A. Sharafeldeen, J. Keowen, and A. Shaffie, "Machine learning approaches for speech-based Alzheimer's detection: A comprehensive survey," *Computers*, vol. 14, no. 2, p. 36, 2025, doi: 10.3390/computers14020036.

[2] X. Qi, Q. Zhou, J. Dong, and W. Bao, "Noninvasive automatic detection of Alzheimer's disease from spontaneous speech: A review," *Frontiers in Aging Neuroscience*, vol. 15, 2023, Art. no. 1224723, doi: 10.3389/fnagi.2023.1224723.

[3] S. de la Fuente Garcia, C. W. Ritchie, and S. Luz, "Artificial intelligence, speech, and language processing approaches to monitoring Alzheimer's disease: A systematic review," *Journal of Alzheimer's Disease*, vol. 78, no. 4, pp. 1547–1574, 2020, doi: 10.3233/JAD-200888.

[4] N. Rezaii *et al*., "Voiceprints of cognitive impairment: Analyzing digital voice for early detection of Alzheimer's and related dementias," *npj Dementia*, vol. 1, p. 35, 2025, doi: 10.1038/s44400-025-00040-0.

[5] J. Yuan, X. Cai, Y. Bian, Z. Ye, and K. Church, "Pauses for detection of Alzheimer's disease," *Frontiers in Computer Science*, vol. 2, 2021, Art. no. 624488, doi: 10.3389/fcomp.2020.624488.

[6] K. Ding, M. Chetty, A. N. Hoshyar, T. Bhattacharya, and B. Klein, "Speech based detection of Alzheimer's disease: A survey of AI techniques, datasets and challenges," *Artificial Intelligence Review*, vol. 57, p. 325, 2024, doi: 10.1007/s10462-024-10961-6.

[7] V. Vincze *et al*., "Telltale silence: Temporal speech parameters discriminate between prodromal dementia and mild Alzheimer's disease," *Journal of Communication Disorders*, vol. 89, 2020, Art. no. 105978.

[8] G. Khatri, R. Soleimani, K. L. Haley, A. Jacks, and E. Lobaton, "Alzheimer's disease classification from speech pause distributions with context information," in *Proc. 46th Annu. Int. Conf. IEEE Engineering in Medicine and Biology Society (EMBC)*, 2024, pp. 1–6, doi: 10.1109/EMBC53108.2024.10782834.

[9] R. A. Sluis *et al*., "An automated approach to examining pausing in the speech of people with dementia," *American Journal of Alzheimer's Disease & Other Dementias*, vol. 35, pp. 1–8, 2020, doi: 10.1177/1533317520939773.

[10] J. Liu *et al*., "Efficient pause extraction and encode strategy for Alzheimer's disease detection using only acoustic features from spontaneous speech," *Brain Sciences*, vol. 13, no. 3, p. 477, 2023, doi: 10.3390/brainsci13030477.

[11] C. Themistocleous, M. Eckerström, and D. Kokkinakis, "Voice quality and speech fluency distinguish individuals with mild cognitive impairment from healthy controls," *PLOS ONE*, vol. 15, no. 7, 2020, Art. no. e0236009, doi: 10.1371/journal.pone.0236009.

[12] N. A. Vats, P. Barche, G. S. Mirishkar, and A. K. Vuppala, "Exploring high spectro-temporal resolution for Alzheimer's dementia detection," in *Proc. IEEE Int. Conf. on Signal Processing and Communications (SPCOM)*, 2022, pp. 1–5, doi: 10.1109/SPCOM55316.2022.9840847.

[13] I. Oiza-Zapata and A. Gallardo-Antolín, "Alzheimer's Disease Detection from Speech Using Shapley Additive Explanations for Feature Selection and Enhanced Interpretability," Electronics, vol. 14, no. 11, p. 2248, 2025, doi: 10.3390/electronics14112248.

[14] K. Ahn *et al*., "Deep learning of speech data for early detection of Alzheimer's disease in the elderly," *Bioengineering*, vol. 10, no. 9, p. 1093, 2023, doi: 10.3390/bioengineering10091093.

[15] **Y. Pu and W.-Q. Zhang,** "Integrating Pause Information with Word Embeddings in Language Models for Alzheimer's Disease Detection from Spontaneous Speech," in *Proc. ICASSP 2025 – IEEE Int. Conf. Acoustics, Speech and Signal Processing (ICASSP)*, 2025, pp. 1–5, doi: 10.1109/ICASSP49660.2025.10888563.

[16] **S. Mohsen,** "Alzheimer's disease detection using deep learning and machine learning: A review," *Artificial Intelligence Review*, vol. **58**, p. **262**, 2025, doi: **10.1007/s10462-025-11258-y**.

[17] **M. Alsuhaibani et al.,** "A review of machine learning approaches for non-invasive cognitive impairment detection," *IEEE Access*, vol. **13**, pp. **56355–56384**, 2025, doi: **10.1109/ACCESS.2025.3555176**.

[18] M. M. Ahsan, S. A. Luna, and Z. Siddique, "Machine-learning-based disease diagnosis: A comprehensive review," *Healthcare*, vol. 10, no. 3, p. 541, 2022, doi: 10.3390/healthcare10030541.

[19] A. S. Alatrany, W. Khan, A. Hussain, H. Kolivand, and D. Al-Jumeily, "An explainable machine learning approach for Alzheimer's disease classification," *Scientific Reports*, vol. 14, p. 2637, 2024, doi: 10.1038/s41598-024-51985-w.

[20] A. C. Mmadumbu, F. Saeed, F. Ghaleb, and S. N. Qasem, "Early detection of Alzheimer's disease using deep learning methods," *Alzheimer's & Dementia*, vol. 21, p. e70175, 2025, doi: 10.1002/alz.70175.

[21] **W. Hechkel and A. Helali,** "Early detection and classification of Alzheimer's disease through data fusion of MRI and DTI images using the YOLOv11 neural network," *Frontiers in Neuroscience*, vol. **19**, Art. no. **1554015**, 2025, doi: **10.3389/fnins.2025.1554015**.

[22] M. Gomis-Pastor *et al*., "Clinical validation of digital healthcare solutions: State of the art, challenges and opportunities," *Healthcare*, vol. 12, no. 11, p. 1057, 2024, doi: 10.3390/healthcare12111057.

[23] A. Bastanfard, M. Fazel, A. A. Kelishami, and M. Aghaahmadi, "The Persian linguistic based audio-visual data corpus, AVA II, considering coarticulation," in *MMM 2010, LNCS 5916*, S. Boll *et al*., Eds. Berlin, Germany: Springer, 2010, pp. 284–294, doi: 10.1007/978-3-642-11301-7_30.

[24] S. Lei, R. Yang, and C.-R. Huang, "Automatic analysis of linguistic features in journal articles of different academic impacts with feature engineering techniques," *arXiv preprint*, arXiv:2111.07525, 2021.

[25] J. Xiao, W. Guo, J. Liu, and M. Li, “Generalization gap in data augmentation: Insights from illumination,” *arXiv preprint*, arXiv:2404.07514v3, 2024.

[26] M. Fathi Ahmadsaraei, A. Bastanfard, and A. Amini, “OBGESS: Automating original Bender Gestalt Test based on one stage deep learning,” *International Journal of Computational Intelligence Systems*, vol. 16, p. 178, 2023, doi: 10.1007/s44196-023-00353-z.

[27] A. Topol, “High-performance medicine: the convergence of human and artificial intelligence,” Nature Medicine, vol. 25, no. 1, pp. 44–56, 2019, doi: 10.1038/s41591-018-0300-7.

[28] F. García-Gutiérrez *et al*., “Unveiling the sound of the cognitive status: Machine learning-based speech analysis in the Alzheimer’s disease spectrum,” *Alzheimer’s Research & Therapy*, vol. 16, p. 26, 2024, doi: 10.1186/s13195-024-01394-y.

[29] C. R. Jack *et al.*, “*Preclinical Alzheimer’s disease: Pathophysiology, biomarker staging, and implications for early detection*,” *medRxiv*, Jan. 2023, doi: 10.1101/2023.01.26.23285064**.**

[30] J. T. Becker, F. Boller, O. L. Lopez, J. Saxton, and K. L. McGonigle, “The natural history of Alzheimer’s disease: Description of study cohort and accuracy of diagnosis,” *Archives of Neurology*, vol. 51, no. 6, pp. 585–594, 1994, doi: 10.1001/archneur.1994.00540180063015**.**

[31] S. Saeedi, S. Hetjens, M. O. W. Grimm, and B. Barsties V. Latoszek, “Acoustic Speech Analysis in Alzheimer’s Disease: A Systematic Review and Meta-Analysis,” J. Prev. Alzheimers Dis., vol. 11, no. –, pp. –, 2024, doi: 10.14283/jpad.2024.132.

[32] R. Soleimani, S. Guo, K. L. Haley, A. Jacks, and E. Lobaton, “The impact of pause and filler word encoding on dementia detection with contrastive learning,” Applied Sciences, vol. 14, no. 19, p. 8879, Oct. 2024, doi: 10.3390/app14198879.

[33] Z. Dong, Z. Zhang, W. Xu, J. Han, J. Ou, and B. W. Schuller, “HAFFormer: A hierarchical attention-free framework for Alzheimer’s disease detection from spontaneous speech,” arXiv:2405.03952 [cs.SD], May 2024.

[34] **Z. Dong, Z. Zhang, W. Xu, J. Han, J. Ou, and B. W. Schuller,** “HAFFormer: A hierarchical attention-free framework for Alzheimer’s disease detection from spontaneous speech,” *in Proc. IEEE Int. Conf. Acoustics, Speech and Signal Processing (ICASSP)*, 2024, pp. 11246–11250, doi: **10.1109/ICASSP48485.2024.10446795**.

[35] V. M. K., G. V., A. P. R., D. S., and S. K. G., “Alzheimer’s disease detection from spontaneous speech and text: A review,” arXiv preprint, arXiv:2307.10005, 2023, doi: 10.48550/arXiv.2307.10005.

[36] **K. Ding et al.,** “Detection of Mild Cognitive Impairment from Non-Semantic, Acoustic Voice Features: The Framingham Heart Study,” *JMIR Aging*, 2024, e55126, doi: **10.2196/55126**.

[37] L. Gauder et al., "The Unreliability of Acoustic Systems in Alzheimer’s Speech Datasets with Heterogeneous Recording Conditions," arXiv:2409.12170v1 [cs.SD], Sep. 2024.

[38] D. He et al., "Automated bimodal pause analysis for acoustic markers of cognitive decline and Alzheimer’s disease in connected speech," Alzheimer’s & Dementia, vol. 21, p. e70635, 2025, doi: 10.1002/alz.70635.

[39] F. García-Gutiérrez et al., "Harnessing acoustic speech parameters to decipher amyloid status in individuals with mild cognitive impairment," Frontiers in Neuroscience, vol. 17, Art. no. 1221401, 2023, doi: 10.3389/fnins.2023.1221401.

[40] A. Bastanfard, A. A. Kelishami, M. Fazel and M. Aghaahmadi, "A comprehensive audio-visual corpus for teaching sound Persian phoneme articulation," 2009 IEEE International Conference on Systems, Man and Cybernetics, San Antonio, TX, USA, 2009, pp. 169-174, doi: 10.1109/ICSMC.2009.5346591.